\documentclass[10pt,prapplied,nofootinbib,notitlepage,twocolumn,superscriptaddress]{revtex4-1}

\usepackage{color,xcolor}

\usepackage{footnote}

\usepackage[utf8]{inputenc}
\usepackage[T1]{fontenc}
\usepackage[english]{babel}  % english language

\usepackage{times}

\usepackage[colorlinks=true,linkcolor=blue,citecolor=blue,urlcolor=blue,pdfauthor={ },pdftitle={ },pdfsubject={ },pdfkeywords={ }]{hyperref}
\usepackage{amssymb,amsmath,amsfonts,amsthm}
\usepackage{mathtools}
\usepackage{verbatim}
\usepackage{enumerate}
\usepackage{graphicx}
\usepackage{hyperref}
\usepackage{bbm}
\usepackage{booktabs}

\usepackage[caption=false]{subfig}

\usepackage{color}
\definecolor{dred}{rgb}{.8,0.2,.2}
\definecolor{ddred}{rgb}{.8,0.5,.5}
\definecolor{dblue}{rgb}{.2,0.2,.8}
\definecolor{dgreen}{rgb}{.2,0.5,.2}

\newcommand{\bra}[1]{\mbox{$\langle #1|$}}
\newcommand{\ket}[1]{\ensuremath{|#1\rangle}}

\newcommand{\be}{\begin{equation}}
\newcommand{\ee}{\end{equation}}
\newcommand{\bea}{\begin{eqnarray}}
\newcommand{\eea}{\end{eqnarray}}

\begin{document}

\title{Quantum Pure State Tomography via Variational Hybrid Quantum-Classical Method}

\author{Tao Xin}
\email{xint@sustech.edu.cn}
\affiliation{Shenzhen Institute for Quantum Science and Engineering and Department of Physics, Southern University of Science and Technology, Shenzhen 518055, China}

\author{Xinfang Nie}
\affiliation{Shenzhen Institute for Quantum Science and Engineering and Department of Physics, Southern University of Science and Technology, Shenzhen 518055, China}

\author{Xiangyu Kong}
\affiliation{State Key Laboratory of Low-dimensional Quantum Physics and Department of Physics, Tsinghua University, Beijing 100084, China}

\author{Jingwei Wen}
\affiliation{State Key Laboratory of Low-dimensional Quantum Physics and Department of Physics, Tsinghua University, Beijing 100084, China}

\author{Dawei Lu}
\email{ludw@sustech.edu.cn}
\affiliation{Shenzhen Institute for Quantum Science and Engineering and Department of Physics, Southern University of Science and Technology, Shenzhen 518055, China}

\author{Jun Li}
\email{lij3@sustech.edu.cn}
\affiliation{Shenzhen Institute for Quantum Science and Engineering and Department of Physics, Southern University of Science and Technology, Shenzhen 518055, China}

\begin{abstract}
To   obtain a complete description of a quantum system, one usually employs standard quantum state tomography, which however   requires exponential number of measurements to perform and hence is  impractical when the system's size grows large. In this work, we introduce a self-learning tomographic scheme   based on the variational hybrid quantum-classical method. The key part of the scheme is a learning  procedure, in which we learn a control sequence capable of   driving the  unknown target state coherently to a simple fiducial  state,  so that the target state can be directly reconstructed by applying the control sequence reversely.  In this manner, the state tomography problem is converted to a state-to-state transfer problem. To solve the latter problem, we use  the closed-loop learning control approach. Our scheme is further experimentally tested  using techniques of a 4-qubit nuclear magnetic resonance.   {Experimental results indicate that the proposed tomographic scheme can handle a broad class of states including entangled states in quantum information, as well as dynamical states of quantum many-body systems common to condensed matter physics.}
%In   experiments, we demonstrate the high-quality and effective  tomographic procedure    of the dynamical state of quantum many-body system and the entangled state using our method.
  \end{abstract}

\maketitle
\section{Introduction}
Quantum state tomography (QST) is the art of determining a quantum state from making   measurements on a set of informationally complete observables \cite{NC00}. It plays a  vital role in many  quantum information processing tasks, such as in characterizing   an interested target quantum system  or in   estimating the performance of a quantum computing experiment. 
However, QST experiments are subject to several  crucial challenges. First,   reconstructing the full   density matrix of a quantum system is highly demanding in the sense that the resources  required  grow exponentially with the system size.  Second, the reliability of QST   is limited by sensitivity to statistical   noise and experimental errors. These   issues impose great difficulties in the practical applications of   QST    even for modest-sized quantum systems \cite{lvovsky2009continuous,baur2012benchmarking,hofheinz2009synthesizing,jullien14quantum}.  It is hence important to develop a robust, precise, and easy-to-implement method for determining unknown quantum states.

Significant efforts have been devoted to improving the performance of QST \cite{cramer2010efficient,gross2010quantum,flammia2011direct,baumgratz2013scalable,ferrie2014self,lanyon2017efficient,chapman2016experimental,ahn2019adaptive},  including self-guided QST, adaptive QST, QST via reduced density matrices (RDMs), and QST via trained neural networks. Self-guided QST considers tomography as a projection measurement optimization problem and finds the optimal solution by stochastic approximation algorithms, which is robust against the noises in  characterizing states. Adaptive QST approach in which the choice of the next measurement depends on the previous measurement outcomes may be as impractical as traditional QST for larger system size \cite{mahler2013adaptive,okamoto2012experimental,huszar2012adaptive}. QST via RDMs measures only the local RDMs to determine the global state so that QST is significantly simplified 
by reducing the measurement resources \cite{xin2017quantum, wyderka2017almost,parashar2009n,chen2013uniqueness}. Measuring   local RDMs are usually convenient  on realistic physical setups. Recent researches show that machine learning methods, e.g.,   multi-layer trained neural work, are  promising   to recover target  states efficiently from the local information  via RDMs \cite{torlai2018neural, gao2017efficient,kieferova2017tomography,gao2018experimental,xin2018local}. Yet, in principle, how to recover an unknown quantum state from its local RDMs is generally an unsolved problem \cite{qi2013quantum}. 

Recently,  there has been a growing interest in the variational hybrid quantum-classical (HQC) approach, which is regarded as a  strategy to boost the efficiency of quantum computational tasks before quantum supremacy is achieved \cite{bauer2016hybrid,bravyi2016trading,mcclean2016theory,wecker2015progress}. In this approach, a quantum computer works in conjunction with classical routines to maximally reduce the requirements for expensive quantum resources. The difficult part of the target problem is accomplished on a quantum computer,   while the relatively easier part is done with a classical computer. HQC approach is a novel attempt versus full-quantum computation \cite{li2017efficient,arrasmith2018variational}. It has found many successful applications ranging from quantum chemistry simulation \cite{kandala2017hardware,peruzzo2014variational}, quantum optimal control \cite{li2017hybrid,mcclean2017hybrid}, and quantum error correction \cite{johnson2017qvector} to quantum state diagonalization \cite{larose2018variational}.

%In this work, we propose an effective QST procedure based on the HQC approach in reconstructing the dynamical states of a quantum many-body system, which is a common and important topic in some famous research works \cite{lanyon2017efficient,torlai2018neural}.  A quantum many-body system starts from the initial state $\ket{\textbf{0}}\bra{\textbf{0}}$ and evolves to the final dynamical state $\rho$ after the polynomial evolution time. 

In this work, we propose an effective QST procedure based on the HQC approach. For an $n$-qubit pure state $\rho$ to be reconstructed,  this procedure attempts to find a unitary process $\mathcal{C}(t)$ that drives the system from $\rho$ to $\ket{\textbf{0}}\bra{\textbf{0}}$ with $\ket{\textbf{0}}=\ket{0}^{\otimes n}$.      Once such a unitary process is found, the unknown state is directly obtained as $\rho=\mathcal{C}^{\dagger}\ket{\textbf{0}}\bra{\textbf{0}}\mathcal{C}$.   {In our framework, $\rho$ is assumed to be a quantum experimental  state, e.g., the final state of a quantum computing experiment. Despite   lacking  information regarding how $\rho$ is realized,   it should come  from a tractable process in the sense that it is prepared from a polynomially-scaled  quantum process. Intractable processes are not expected to be observed  or realized in experiment, thus only low complexity states are experimentally accessible \cite{Fernando2019}. This implies that the optimal unitary trajectory that  connects $\rho$ and $\ket{\textbf{0}}=\ket{0}^{\otimes n}$ would not be unreasonably long.  } The unitary trajectory is realized through a parameterized controlled evolution $\mathcal{C}(b,t)$, where $b$ represents the set of control parameters. So, our procedure actually seeks an optimal control sequence. In this way, the state tomography problem converts to an optimal control problem. In searching for the optimal control sequence, the  target function to be optimized is chosen to be the distance between   $\ket{\textbf{0}}\bra{\textbf{0}}$ and the actual final state. This distance can be easily estimated with a few measurements for many experimental platforms. In our construction, a trusted quantum computer replaces a classical computer to efficiently perform the evolution of the sequence $\mathcal{C}(b,t)$ and precisely estimate both  target function and its gradient in each iteration.  Classical computer takes charge of determining the search direction and the step size to update the control parameters $b$. 

% Prototypical examples include entangled states prepared via a quantum circuit of polynomial size and dynamical states of locally interacting quantum many-body systems.

QST using the HQC approach is a promising and applicable
technique for reconstructing quantum states in current physical
setups. Here, we also present an experimental demonstration of QST via HQC on a 4-qubit NMR quantum processor.   {We basically consider two types of quantum states, namely dynamical states of quantum many-body systems and entangled states. In the experiment, we successfully drove such states to the ground state, and reconstructed them at high-quality, without involving performing informationally-complete measurements. Therefore, the feasibility and the ability of our proposed HQC based-QST method has been confirmed.}
% We successfully drove a 4-qubit dynamical state of quantum many-body system and an entangled state to the ground state, and reconstructed the target states with the high-quality in experiments.  The feasibility and the ability of self-learning QST in reconstructing quantum states are fully demonstrated.}

\begin{figure*}
%\centering  
\includegraphics[width=0.75\linewidth]{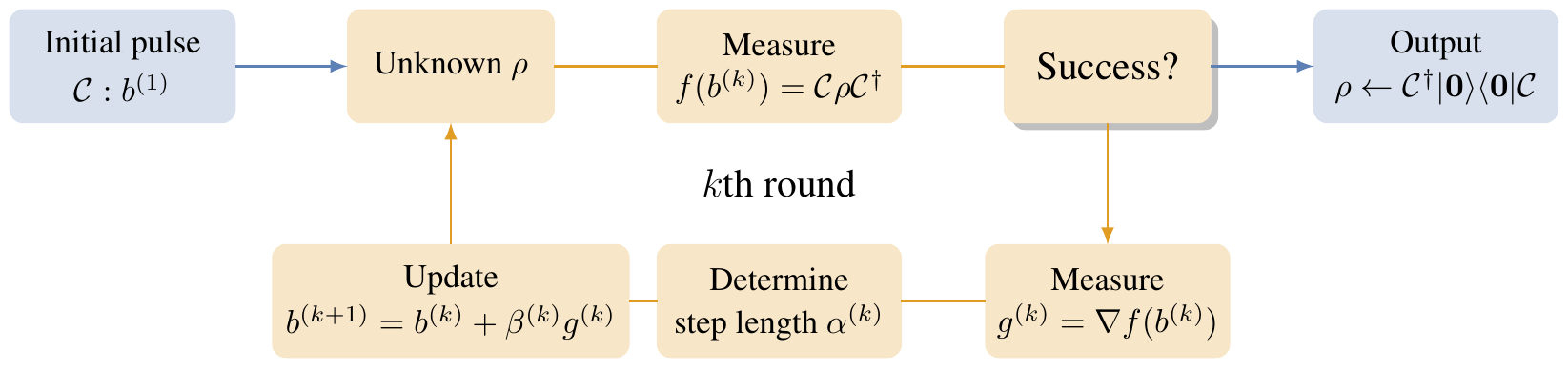}  
\caption{The workflow and schematic diagram of QST via HQC approach. The reconstruction of an unknown state $\rho$ is iteratively finished based on gradient-based searching. For each feed pulse $b^{(q)}$, the fitness function $f(b^{(q)})$ and  its gradient   $g(b^{(q)})=\nabla f(b^{(q)})$ are computed by the system itself,  while classical computer takes charge of the storage and update of the control pulse $b^{(q)}$ according to the   $f$ and $g$.    When the function $f$ reaches the desired accuracy, the optimal control pulse $b$ is obtained and  is used to reconstruct $\rho$.} 
\label{process}
\end{figure*}

\section{Method} 
%Parameterized quantum circuits (PQCs) have been broadly used as a hybrid quantum-classical   learning scheme to accomplish generative tasks.
%It is assumed that one can prepare states belonging to some parameterized family $\rho_{\bm{x}}$ for $\bm{x} = (x_1, \ldots, x_M) \in \mathcal{X} \subset \mathbb{R}^M$, where $M$ is the number of variational
%parameters. The set of parameterized states which may be prepared will depend on the specifications
%of the quantum device. We will consider parameterizations consisting of $M$  pulses  applied to some easy-to-prepare starting state $\rho_i$:
%\begin{align}
%	\rho_{\bm{x}} & = U(\bm{x}) \rho_i U(\bm{x})^\dag  \nonumber \\
%	& = U(x_M) \cdots U(x_1)  \rho_i U(x_1)^\dag \cdots U(x_M)^\dag.  \nonumber
%\end{align}
%where   is the Hermitian operator which generates pulse $j$. Here $\rho_i = |\bm{0}\rangle \langle \bm{0}|$ is the all-zeros starting qubit state for $n$ qubits.

\subsection{Problem Setting}
To start, we first introduce the problem by describing the general control setting that we address. We consider an $n$-qubit   spin system with   internal Hamiltonian $\mathcal{H}_0$ and   control Hamiltonian $\mathcal{H}_c$. Normally, $\mathcal{H}_0$ is a  two-body local Hamiltonian with constant system parameters. $\mathcal{H}_0$ together with $\mathcal{H}_c$ provide the ability to engineer the system with full controllability. The control is realized by a time-dependent   magnetic field, namely, $\mathcal{H}_c=\sum^n_{i=1}\{b_x(t)\sigma^i_x+b_y(t)\sigma^i_y\}$, with $\sigma_x$, $\sigma_y$, and $\sigma_z$ being the three Pauli matrices, and $b(t)=\{b_x(t), b_y(t)\}$ being the control sequence to manipulate the dynamics of the quantum system.
Consider the situation when the spin system processor finishes some quantum task  we submit, the final state $\rho$ contains the useful information to be extracted, or when there is an unknown state $\rho$ for which we are unaware of its earlier evolution history. The problem is to reconstruct $\rho$.%The commonly-known technique to achieve this goal is to perform full QST: measure the exponentially many coefficients of the density matrix $\rho$. 

The conventional QST method works as follows. For an  $n$-qubit unknown quantum state $\rho$, one can decompose it in terms of the   Pauli product operator basis as $\rho=\sum^3_{i_1,..,i_n=0}\gamma_{i_1,..,i_n}\sigma_{i_1}\otimes \cdots \otimes \sigma_{i_n}$. Here, $\sigma_0=I$, $\sigma_1=\sigma_x$, $\sigma_2=\sigma_y$, and $\sigma_3=\sigma_z$. The coefficient $\gamma_{i_1,..,i_n}$ is the projective component of $\rho$ in   $\sigma_{i_1}\otimes \cdots \otimes \sigma_{i_n}$. The standard QST  estimates all the coefficients  from a series of quantum measurements. There are in total $4^n-1$ and $2^n-1$  coefficients to be  determined for   mixed states and pure states, respectively. It is obvious that the exponential growth in the number of experiments needed to measure all of these coefficients  results in the difficulty in performing standard QST for large quantum systems.
 
\subsection{Our Variational HQC Approach}
Here, we show that, instead of using traditional QST,   the state $\rho$ can be reconstructed via a variational learning procedure in an iterative way. We first search an optimal control sequence $b(t)$ to drive the unknown state $\rho$ to the ground state $\ket{\textbf{0}}\bra{\textbf{0}}$ and then realize the reconstruction task simply as $\rho=\mathcal{C}(b)^{\dagger}\ket{\textbf{0}}\bra{\textbf{0}}\mathcal{C}(b)$. We have to choose a reasonable fitness function $f(b)$ as a function of the parameters $b$, to evaluate the performance of  $\mathcal{C}(b)$. The function is defined as: $f(b)=D(\mathcal{C}(b)\rho\mathcal{C}(b)^{\dagger},\ket{\textbf{0}}\bra{\textbf{0}})=\text{Tr}(\ket{\textbf{0}}\bra{\textbf{0}} \cdot \mathcal{C}(b)\rho\mathcal{C}(b)^{\dagger})$. The physical picture behind $f(b)$ is that it measures the overlap between the final state and the state $\ket{\textbf{0}}\bra{\textbf{0}}$. This value can be easily obtained with applying the projective measurement operator $\ket{\textbf{0}}\bra{\textbf{0}}$ in the computational basis.

Now, the key task is to solve   a   constrained optimization problem, that is, to find $b(t)$,
\begin{align}
\text{max } & \quad  {}  f(b)=\text{Tr}(\ket{\textbf{0}}\bra{\textbf{0}} \cdot \mathcal{C}(b)\rho\mathcal{C}(b)^{\dagger}), \nonumber \\ 
\text{s.t. }  & \quad {} \mathcal{\dot C}(b)=-i(\mathcal{H}_0+\mathcal{H}_c(b)) \mathcal{C}(b).   \nonumber
\end{align}
It is noted that, while this is a standard state-to-state optimal control problem, it can not be solved on a classical computer because   the state $\rho$ is unknown. Besides,  simulating the system's dynamics under $b(t)$ can be infeasible for a large  quantum system. To circumvent the difficulty, we utilize the fact that the process of optimizing $\mathcal{C}(b)$ including the computing of the  target function $f(b)$ and its gradient $g(b)=\nabla f(b) $ can be done with   the controlled spin system itself \cite{li2017hybrid}. On the other hand, a classical computer collects the gradient information from the experiments, stores the control parameters, determines the search direction, generates the next control sequence and feeds the iterate into the target system until the desired termination condition is fulfilled. As a result, such a QST process forms a closed loop, as   illustrated in Fig. \ref{process}.  More concretely, QST via HQC is divided into the following steps:

(i) For  numerical optimization, we discretize the control sequence $b(t)$ by dividing it into $M$ slices
\begin{align}
b(t)=\{b_x[1], b_x[2], ..., b_x[M], b_y[1], b_y[2], ..., b_y[M]\}.
\end{align}
The time length of each slice  is a constant $\tau=T/M$ and the amplitude $b_{x, y}[m]$ in the $m$-th slice is also a constant. Then the   propagator of the $m$-th slice is
\begin{equation}
	\mathcal{C}_m=\text{exp}\left\{-i\tau(\mathcal{H}_0+b_x[m]\sum^n_{i=1}\sigma^i_x+b_y[m]\sum^n_{i=1}\sigma^i_y)\right\}. 
\end{equation}
The total evolution of the sequence $b$ with $M$ slices can be described as $\mathcal{C}=\mathcal{C}_M \mathcal{C}_{M-1} \cdots \mathcal{C}_2\mathcal{C}_1$. Hence, the  target function can be written as,
\begin{align}
f(b)=\text{Tr}(\ket{\textbf{0}}\bra{\textbf{0}} \cdot \mathcal{C}_M \cdots\mathcal{C}_1\rho\mathcal{C}_1^{\dagger}\cdots  \mathcal{C}_M^{\dagger}).
\label{ff}
\end{align}

(ii) A  randomly generated set of pulse parameters $b^{(0)}_x[m]$ and $b^{(0)}_y[m]$ is chosen as the initial guess.  Now we will describe two    methods \cite{li2017hybrid,mcclean2016theory}    to calculate the gradient values $g_{x,y}[m]$. 

\emph{Method 1.}--As long as the duration $\tau$ is small enough, the gradient value of the $m$-th slice $g_{\alpha}[m]=\partial f/\partial b_{\alpha}[m]$ with $\alpha=x,y$ can be approximately computed as
\begin{align}
g_{\alpha}[m]=\sum^n_{i=1}\text{Tr}(-i\tau\ket{\textbf{0}}\bra{\textbf{0}} \cdot \mathcal{C}^{M}_{m+1}[\sigma^i_{\alpha}, \mathcal{C}^{m}_{1}\rho\mathcal{C}^{m \dagger}_{1}] \mathcal{C}^{M \dagger}_{m+1}).   
\end{align}
Here, $\mathcal{C}^{M}_{m+1}=\mathcal{C}_M \cdots \mathcal{C}_{m+2}\mathcal{C}_{m+1}$ and $\mathcal{C}^{m}_{1}=\mathcal{C}_m  \cdots \mathcal{C}_2\mathcal{C}_1$. For any state $\rho$, one can check that there is
\begin{align}
[\sigma^i_{\alpha}, \rho]=i[\mathcal{R}^i_{\alpha}(\frac{\pi}{2})\rho\mathcal{R}^i_{\alpha}(\frac{\pi}{2})^{\dagger}-\mathcal{R}^i_{\alpha}(-\frac{\pi}{2})\rho\mathcal{R}^i_{\alpha}(-\frac{\pi}{2})^{\dagger}],  \nonumber
\end{align}
where $\mathcal{R}^i_{\alpha}(\pm \pi/2)$ is the $\pm \pi/2$ rotation around $\alpha$ axis acting on the $i$-th qubit. If we define the notations $\mathcal{C}^{im}_{\pm\alpha}=\mathcal{C}_M...\mathcal{C}_{m+1}\mathcal{R}^i_{\alpha}(\pm \pi/2)\mathcal{C}_{m}...\mathcal{C}_{2}\mathcal{C}_1$, we can obtain the expression for   $g_{\alpha}[m]$: 
\begin{align}
\tau\sum^n_{i=1}\left\{\text{Tr}(\ket{\textbf{0}}\bra{\textbf{0}} \cdot \mathcal{C}^{im}_{+\alpha}\rho \mathcal{C}^{im \dagger}_{+\alpha})
-\text{Tr}(\ket{\textbf{0}}\bra{\textbf{0}} \cdot \mathcal{C}^{im}_{-\alpha}\rho \mathcal{C}^{im \dagger}_{-\alpha}) \right\}. \nonumber
\end{align}
The two terms in the above equation    are similar to that in  the  target function $f(b)$ in Eq. (\ref{ff}), because $\mathcal{C}^{im}_{\pm\alpha}$ is created by simply inserting a local operation $\mathcal{R}^i_{\alpha}(\pm \pi/2)$ between the $m$-th and the ($m+1$)-th slice  in $b(t)$. Hence, the $m$-th gradient $g_{\alpha}[m]$ can be   obtained from a quantum system itself by performing $2n$ measurement experiments. In total, we   need one experiment for measuring the  target function $f$ and additionally $4nM$ experiments for measuring the $2M$-dimensional gradient vector $g$. The number of required experiments is thus linear with the number of qubits.\\

\emph{Method 2.}-- {Finite-difference approximation.} To estimate the gradient value of the $m$-th slice $g_{\alpha}[m]$, we directly change the $m$-th control parameter $b_{\alpha}[m]$ by a step size $\delta$ in forward direction and create a new control sequence $b: b_{\alpha}[m]+ \delta$. We apply the  sequence  to the controlled system and then measure its corresponding  target function value  $f(b: b_{\alpha}[m] + \delta)$ in the same way as measuring $f(b)$. In the first-order approximation, the gradient value of the $m$-th slices $g_{\alpha}[m]$ can be written as
\begin{align}
g_{\alpha}[m]=\frac{f(b: b_{\alpha}[m] +\delta)-f(b: b_{\alpha}[m])}{\delta}.
\end{align}
The step size $\delta$ should be cautiously chosen to guarantee   convergence. Usually, we can fix $\delta$ as a sufficiently small value compared with the range of $b$. In this method, we only need $2M+1$ experiments to  determine the gradient values and the target function, independent with the number of qubits.

It should be noted that for both methods, the value of $\tau$ has to be kept small. As such, if the  coupling strengths between the different spins are larger, then the shorter total time of the control sequence is required, and accordingly smaller   number of the slices $M$ would suffice. 

(iii) Next,   we   determine the search direction and generate the control sequence for the next iteration. Suppose that the control sequence in the $k$-th iteration is $b^{(k)}=\{b^{(k)}_x[m], b^{(k)}_y[m]\}$ $(m=1, ..., M)$ and the measured gradient   is $g^{(k)}=\{g^{(k)}_x[m], g^{(k)}_y[m]\}$. Then we can move along the search direction to the next iteration,
\begin{align}
b^{(k+1)}_{\alpha}[m]=b^{(k)}_{\alpha}[m]+\beta_\alpha^{(k)}\cdot g^{(k)}_{\alpha}[m].
\end{align}
We choose an appropriate step length $\beta^{(k)}$ to achieve the optimal increase of $f$ in the gradient direction.  The steps (ii) and (iii) are repeated until the target function $f(b)$ achieves the desired goal or converges to a local extremum.

(iv) With the optimal sequence   at hand,  we thus obtain a representation of the unknown state $\rho$ in terms of a parameterized control sequence. This is in analogy with that in variational quantum algorithms, a quantum state is expressed in terms of a parameterized quantum circuit. In some circumstances,  one can directly   calculate out $\rho = \mathcal{C}(b)^{\dagger}\ket{\textbf{0}}\bra{\textbf{0}}\mathcal{C}(b)$ efficiently. For example, when the system Hamiltonian is one-dimensional, it is possible to simulate the dynamics $\mathcal{C}(b)^{\dagger}$ with high accuracy by means of tensor-network based   techniques like the time-dependent density matrix renormalization group \cite{PhysRevLett.93.076401,RevModPhys.77.259}.   Our method thus provides the possibility to perform efficient QST for tasks in near-term quantum systems.

%What is unideal   for prospects of optimization is that, the resulting classical   optimization problem will generally be complicated and nonconvex, and hence may be intractable. However, one can hope for heuristics which find a reasonable approximate solution.

\subsection{Applications} 
  {In exploiting the potential  applications of our proposed HQC method, one important problem is the scaling issue.} 
Generally, due to the intrinsic complexity of the state tomography problem, it is unlikely to have a tomographic scheme with favorable scaling for any quantum state from the  Hilbert space. What one can do in practice is to presume  a certain class of quantum states, and then to make a  tomographic scheme feasible so that it can   replace the inefficient full state tomography approach. This is meaningful because the states involved in  common experiments do not spread in the whole Hilbert space.   {Actually, in principle most quantum states are exponentially hard to reach or even to approximate, so  most states are in fact beyond the grasp of the quantum experimenters \cite{NC00,PhysRevLett.106.170501}.} Therefore,   the target states to be reconstructed should only locate in a part of the Hilbert space. The scaling issue of our HQC method has to be comprehended under this general picture. That is, without any restriction of the target state, the number of slices required for a control sequence to drive it to $\ket{\textbf{0}}\bra{\textbf{0}}$ tends to be exponentially large. The actual problem of   interest  is to find out whether HQC provides better scaling   for a certain   class of  quantum states.

\begin{figure}[b]
\centering  
\includegraphics[width=0.95\linewidth]{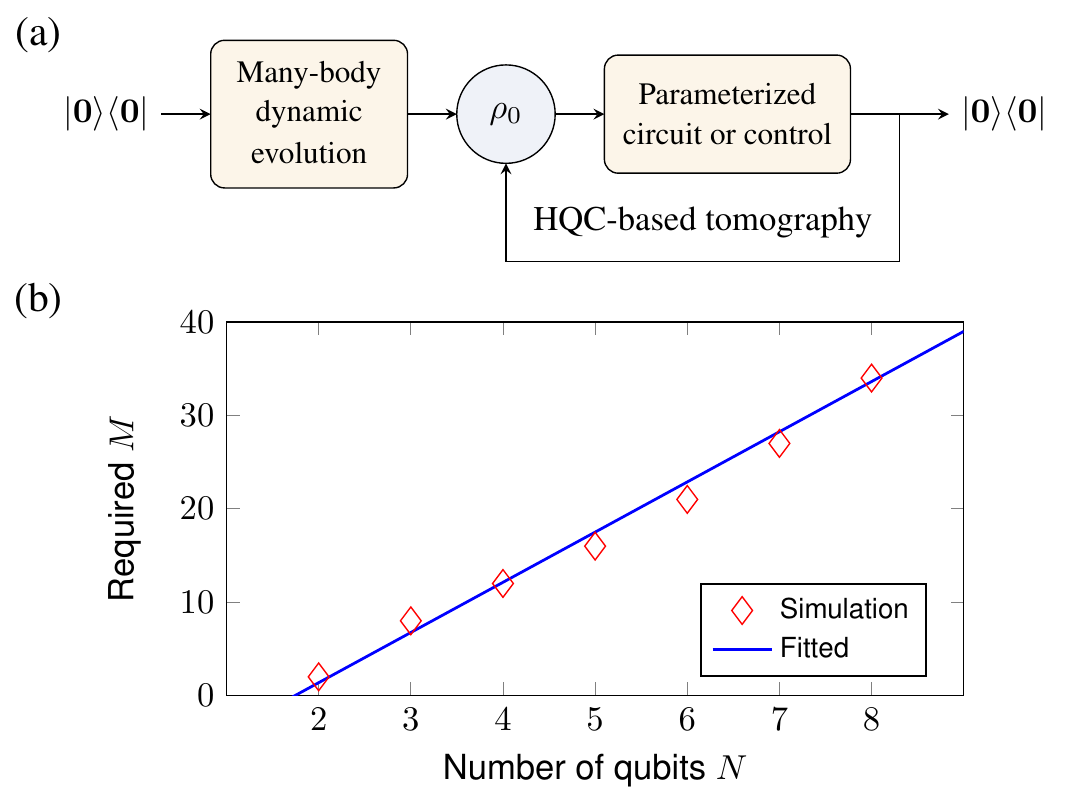}  
\caption{(a) Tomography of a dynamical state of a many-body system using the HQC  approach. (b)  The number of   slices $M$ required against the size of the system. The target state is a time-evolved state at some time $T$: $\rho_0=\ket{\phi(T)}\bra{\phi(T)}$ with $\ket{\phi(T)}=e^{-i\mathcal{H_{\text{0}}}T}\ket{\textbf{0}}$.     In the   simulation, we set $T=3$ s and the size of the system ranges from 2 to 8, and then we seek the minimum number of  slices required for   the target state  $\ket{\phi(T)}$ to be driven towards $\ket{\textbf{0}}\bra{\textbf{0}}$ with  fidelity  over 99\%.    The line is the fitting result over these points.} 
\label{scale}
\end{figure}

  {As an applicative example, we show that the QST via HQC approach can reconstruct dynamical states of quantum many-body systems. Consider the scenario   shown in Fig.  \ref{scale}(a) where we attempt to tomography a dynamical state of a quantum many-body system. There have been   remarkable theoretical findings by researchers in many-body physics and quantum information theory showing that,  dynamical states of   quantum many-body systems  can usually be described by  only a polynomial number of parameters \cite{GVWC07}. It is reasonable to assume that,   such states have  polynomially scaled complexity, even for the case of a chaotic many-body Hamiltonian.  For example, a latest work has proved that the complexity of the quantum state generated by a local random quantum circuit grows linearly   for a long time  \cite{Fernando2019}. Previously it has been demonstrated that, to estimate   dynamical  states of quantum many-body systems,  the machine learning approach can be a simple substitute for full state tomography \cite{Lanyon17,Torlai18}. Here, we   numerically simulate the   potential of using our method for     reconstructing   the dynamical states of  a quantum many-body system and then present an experimental test in the next section. }

%\begin{figure*}
%\centering  
%\includegraphics[width=0.9\linewidth]{scaling.pdf}  
%\caption{  {Numerically found  number of  slices $M$ required against the size of the system, for the following three cases: (a)   Dynamical states of Ising  many-body system with Hamiltonian  Eq. (\ref{Ising}). The target state is a time-evolved state at   time $T$: $\rho=\ket{\phi(T)}\bra{\phi(T)}$ with $\ket{\phi(T)}=e^{-i\mathcal{H_{\text{0}}}T}\ket{\textbf{0}}$.     In the   simulation, we set $T=3$ s and the size of the system ranges from 2 to 8, and then we seek the minimum number of  slices required for   the target state  $\ket{\phi(T)}$ to be driven towards $\ket{\textbf{0}}\bra{\textbf{0}}$ with  fidelity  over 99\%.    The line is the fitting result over these points; (b) Entangled quantum states. In simulation, we choose the following form of   entangled states: $(\ket{0}^{\otimes n}+\ket{1}^{\otimes n})/\sqrt{2}$; (c) The ground states of the 2-local Hamiltonian $\mathcal{H}_\text{Ising}$ in Eq. (\ref{Ising}). }} 
%\label{DKL}
%\end{figure*}

To specify the problem, we consider a quantum many-body system which starts from the initial state $\ket{\textbf{0}}\bra{\textbf{0}}$ and evolves to the final dynamical state $\rho_0$ after a polynomial evolution time $t$. Suppose that just the state $\rho$ is given to us, with no more information of the Hamiltonian and $t$. Despite this, we can be confident about the existence of an efficient  sequence capable of realizing the state transfer  between $\ket{\textbf{0}}\bra{\textbf{0}}$ and $\rho$. For example, the reverse of the transfer can be realized  simply by reversing the many-body Hamiltonian, which can be simulated via a parameterized quantum circuit or a control sequence.   Consequently, although the actual parameters found via our hybrid optimization method may not be exactly the same as the said reverse evolution parameters,   the number of parameters can keep   polynomially scaled with the size of system. To support the statement, here we  choose the reconstruction of  the dynamical states of the Ising-model Hamiltonian as an illustrative example and numerically simulate how the number of parameters required scales with the size of the system using our  HQC approach. In our simulation, the  Ising  Hamiltonian is  
\begin{equation}
	\mathcal{H_{\text{Ising}}}=-\sum_{i=1}^{n-1}\sigma_i^z\sigma_{i+1}^z+\sum_{i=1}^{n}\sigma^x_i,
	\label{Ising}
\end{equation}
and the control Hamiltonian takes the form $\sum_{i=1}^{n}h_i^{(m)}\sigma^x_i+\sum_{i=1}^{n}b_i^{(m)}\sigma^y_i$, where $h_i^{(m)}$ and $b_i^{(m)}$ are the control parameters of the $m$-th slice on the site $i$. From the results in Fig. \ref{scale}(b), it is evident that the  HQC approach scales well with the size of the system. 
%\deleted{Besides, we adopt the same simulation method as that of the dynamical state and also simulate the scalability behavior of reconstructing the entangled quantum states and the ground states of $k$-local Hamiltonian. Their scaling is presented in Fig. \ref{DKL}(b) and Fig. \ref{DKL}(c), respectively.}

\section{Experiment} 

\subsection{System}
As a proof-of-principle demonstration, we experimentally test the feasibility of our method by reconstructing two kinds of quantum states on a 4-qubit NMR simulator \cite{xin2018nuclear,luo2018experimentally,yao2017quantum}. As shown in Fig. \ref{sample}(a), the spins we used are carbon nuclei in $^{13}$C-labeled trans-crotonic acid dissolved in d6-acetone after decoupling them from   the methyl group M and the hydrogen atoms. 
%We drove the spins C$_2$ and C$_3$ to an entangled state $(\ket{00}+\ket {11})/\sqrt{2}$ and characterized it by our method, with the coupling of 72.36 Hz between C$_2$ and C$_3$. 
Our experiments were  carried out on a Bruker 600 MHz spectrometer at room temperature 298 K. Under the weak coupling approximation, the Hamiltonian of our system in reference frame can be written as,
\begin{align}\label{Hamiltonian}
\mathcal{H}_{\text{int}}=\sum\limits_{j=1}^4 {\pi \nu _j} \sigma_z^j  + \sum\limits_{j < k,=1}^4 {\frac{\pi}{2}} J_{jk} \sigma_z^j \sigma_z^k,
\end{align}
with the chemical shifts $\nu_j$ and the J-coupling strengths $\emph{J}_{jk}$, respectively. Figure \ref{sample}(a) gives the  molecular structure and the physical parameters.

\subsection{Scheme}
The experimental procedure can be divided into four parts: (i) Initialize the spins into the initial state $\ket{0000}$; (ii) Prepare the target state $\rho$ from $\ket{0000}$; (iii) At the $k$th iteration, feed the   pulse $b^{(k)}$ into the spin system and measure the fitness function and its gradient; (iv) Update the pulse for the next iteration and perform the iterations until achieving the goal.

First, we initialized the NMR system to a pseudo-pure state (PPS) from the thermal equilibrium state which is a highly mixed state. This was  implemented using the spatial averaging technique involving  unitary operations and gradient fields \cite{cory1998nuclear,cory1997ensemble}. The density matrix of PPS is of the form $\rho_{\text{g}}=(1-\epsilon)\mathbb{I}/16+\epsilon\ket{0000}\bra{0000}$, here $\epsilon\approx10^{-5}$ represents the polarization  and $\mathbb{I}$   the $16\times 16$ identity operator. In the following, we just use the part $\ket{0000}\bra{0000}$ (PPS) as the description of the spins, while ignoring the identity part, because the identity part of the state does not evolve nor contribute to the NMR signals under any unitary operations \cite{knill1998power}. 
Our experimentally  prepared PPS achieved over 0.99 fidelity according to the state fidelity definition $F(\rho,\sigma)=\text{Tr}(\rho\sigma)/\sqrt{\text{Tr}(\rho^2)}\sqrt{\text{Tr}(\sigma^2)}$ \cite{lee2002quantum,leskowitz2004state}. 

Second, we prepared the following target states and then reconstructed them via   HQC-based QST.

\begin{figure}[b]
\centering  
\includegraphics[width=1\linewidth]{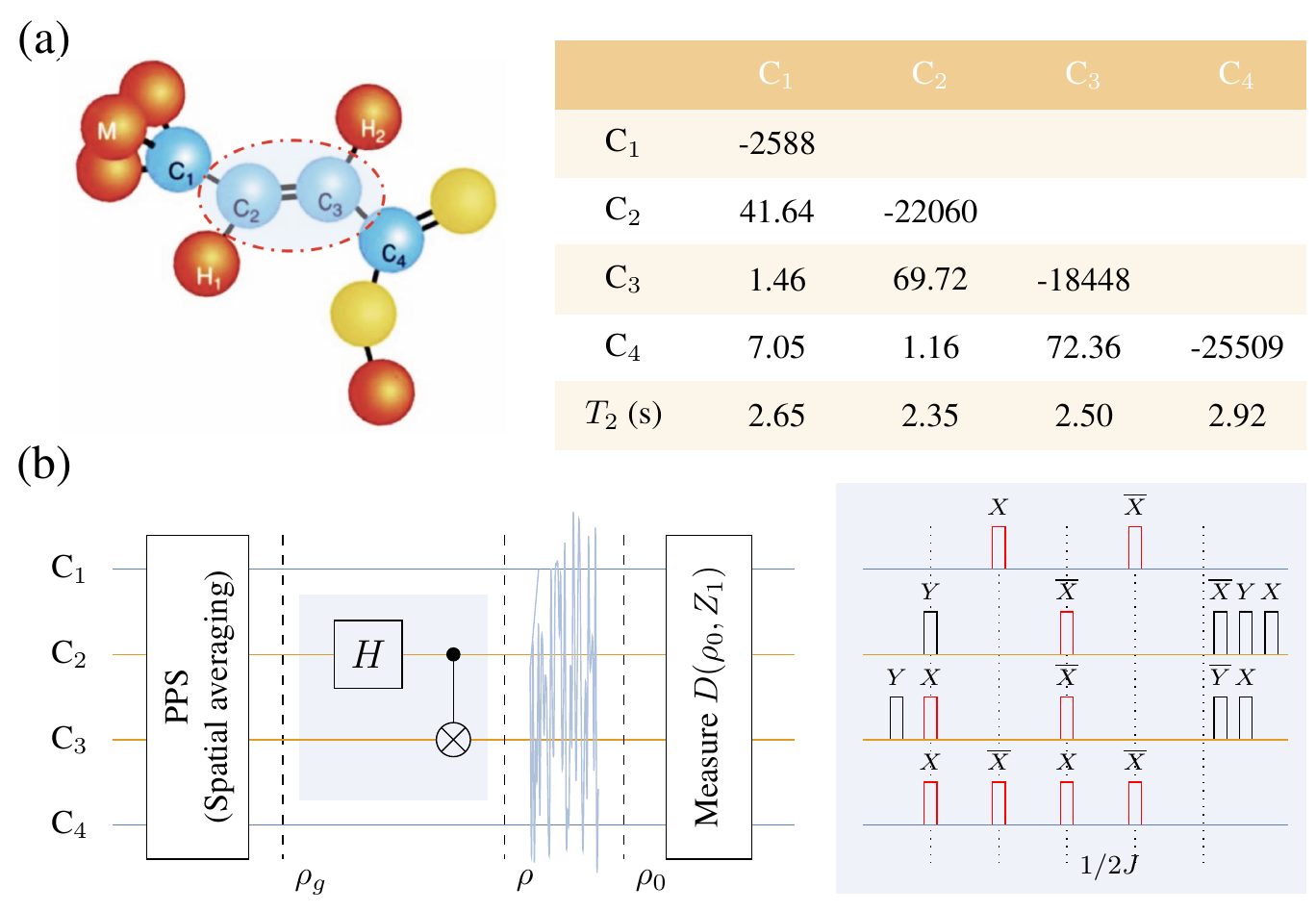}  
\caption{Molecular parameters and experimental quantum circuit. (a) Molecular structure and Hamiltonian parameters for $^{13}$C-labeled trans-crotonic acid. The table shows the chemical shifts and J-couplings in the diagonal and off-diagonal elements, respectively. C$_2$ and C$_3$ have the maximum coupling  value of 72.36 Hz. (b) Quantum circuit for testing the proposed method and NMR pulse sequence for preparing $\rho$. It includes three parts: PPS initialization, the preparation for the target state $\rho$, and the pulse optimization by quantum system itself. Hadamard and CNOT gates can also be realized by a pulse sequence including local rotations and J-coupling evolutions illustrated in the right  of (b), where the  black and red rectangles represent $\pi/2$ and $\pi$ pulses around the directions indicated on top of them, respectively.}
\label{sample}
\end{figure}

  {\emph{(1) Dynamical state of a quantum many-body system}. We choose to   reconstruct  the dynamical states of a prototypical quantum many-body system, namely the transverse field Ising-model. As an illustrative example, we consider the   Hamiltonian $\mathcal{H_{\text{Ising}}}$ in Eq. (\ref{Ising}). The target state is a time-evolved state   $\ket{\phi(T)}=e^{-i\mathcal{H_{\text{Ising}}}T}\ket{\textbf{0}}$ at time $T=0.6~\text{s}$.  In experiment, to prepare $\ket{\phi(T)}$, a 3 ms optimized radio-frequency (RF) pulse with   fidelity 0.995 is applied to the system. This shaped pulse has been designed to be robust to RF inhomogeneity.}

\emph{(2) Entangled quantum state}. We prepare the spins C$_2$ and C$_3$ to an entangled state $(\ket{00}+\ket {11})/\sqrt{2}$ and keep the other spins stay in the $\ket{0}$ state. This can be done via a quantum circuit involving a Hadamard gate   on C$_2$ and a controlled-NOT gate CNOT$_{2,3}$.  In NMR, the Hadamard gate   can be decomposed as $\mathcal{R}^2_x(\pi)\mathcal{R}^2_y(\pi/2)$ and CNOT$_{2,3}$ can be decomposed as
\begin{align}
\text{CNOT}_{2,3} = \sqrt{i}\mathcal{R}^2_z(\frac{\pi}{2})\mathcal{R}^3_z(-\frac{\pi}{2})\mathcal{R}^3_x(\frac{\pi}{2})\mathcal{U}(\frac{1}{2J})\mathcal{R}^3_y(\frac{\pi}{2}),
\label{decompose}
\end{align}
Here, $\mathcal{R}^2_z(\pi/2)$ and $\mathcal{R}^3_z(-\pi/2)$ can be written as
  \begin{align}
\mathcal{R}^2_z(\frac{\pi}{2})&=\mathcal{R}^2_x(\frac{\pi}{2})\mathcal{R}^2_y(\frac{\pi}{2})\mathcal{R}^2_x(-\frac{\pi}{2}), \nonumber \\
\mathcal{R}^3_z(-\frac{\pi}{2})&=\mathcal{R}^3_x(\frac{\pi}{2})\mathcal{R}^3_y(-\frac{\pi}{2})\mathcal{R}^3_x(-\frac{\pi}{2}).  \nonumber
\label{decompose}
\end{align}
The   evolution operator $\mathcal{U}(1/2J)$ represents the $J$-coupling evolution $e^{-i\pi\sigma^2_z\sigma^3_z/4}$ and it  can be realized by inserting refocusing pulses,
\begin{align}
&\mathcal{R}^{2,3,4}_x(\pi) \rightarrow f(\frac{1}{8J}) \rightarrow \mathcal{R}^{1}_x(\pi)\mathcal{R}^{4}_x(-\pi) \rightarrow f(\frac{1}{8J})    \\
& \rightarrow  \mathcal{R}^{2,3}_x(-\pi)\mathcal{R}^{4}_x(\pi) \rightarrow  f(\frac{1}{8J}) \rightarrow \mathcal{R}^{1,4}_x(-\pi) \rightarrow f(\frac{1}{8J}), \nonumber
\label{decompose}
\end{align}
where $f(1/8J)$ represents  the system free evolution of duration $1/8J$,  and $\mathcal{R}^i_\alpha(\theta)$ is a rotation about the axis $\alpha =x$ or $y$ with  angle $\theta$ acting on the $i$-th qubit. The  circuit and the corresponding pulse sequence is  shown in  Fig. \ref{sample}(b). 
%Considering that selective excitations are usually imperfect in homonuclear systems, we packed up all the pulses together and optimized them by the gradient ascent pulse engineering (GRAPE) technique \cite{khaneja2005optimal,ryan2008liquid,moussa2012practical}. They are designed to be insensitive to the inhomogeneity of the rf field and static magnetic field. The GRAPE technique provides the 0.9 ms and 15 ms shaped-pulses with over $99.5\%$ fidelity for realizing $H$ and CNOT gates, respectively.  

After the above state preparation step, a 4-qubit full QST was implemented. The QST results show that the  prepared dynamical state and the entangled state has the experimental fidelity of 0.99 and 0.98, respectively. The purpose of doing the conventional full   QST is for verifying that the target state has indeed been prepared and also for the subsequent comparison with   our method. The basic principle of QST in NMR is as follows \cite{PhysRevA.96.032307}.  NMR measurement is essentially ensemble measurement, one can determine the expectation value of an observable using one experiment provided that the signal-to-noise ratio is good.  The NMR setup   measures the expectation values of the single-quantum coherence operators. If one wants to measure other operators, readout pulses are needed before data acquisition to transfer them    to   detectable single-quantum coherences. Finally, all the measurement data are collected to infer the coefficients of   $\rho$.

Next, following our proposed HQC method, we iteratively optimize a control sequence $b(t)$ to drive the target state towards $\ket{0000}\bra{0000}$.  We start  from an initial guess $b^{(1)}=\{b^{(1)}_x, b^{(1)}_y\}$. When it proceeds to the $k$-th iteration we fed the optimized pulse $b^{(k)}$ into the NMR system, and then estimated the resulting fitness function $D(\rho_0^{(k)}, Z_1)$ from measuring the expectation value of the operator $Z_1=\sigma_z\otimes\ket{000}\bra{000}$. Notice that, optimizing $Z_1$   is   equivalent to optimizing $D(\rho_0^{(k)}, \ket{0000}\bra{0000})$, and   importantly, it can be done with just implementing a single $\pi/2$ readout pulse on the first spin C$_1$.  

\begin{figure*}
\centering  
\includegraphics[width=0.98\linewidth]{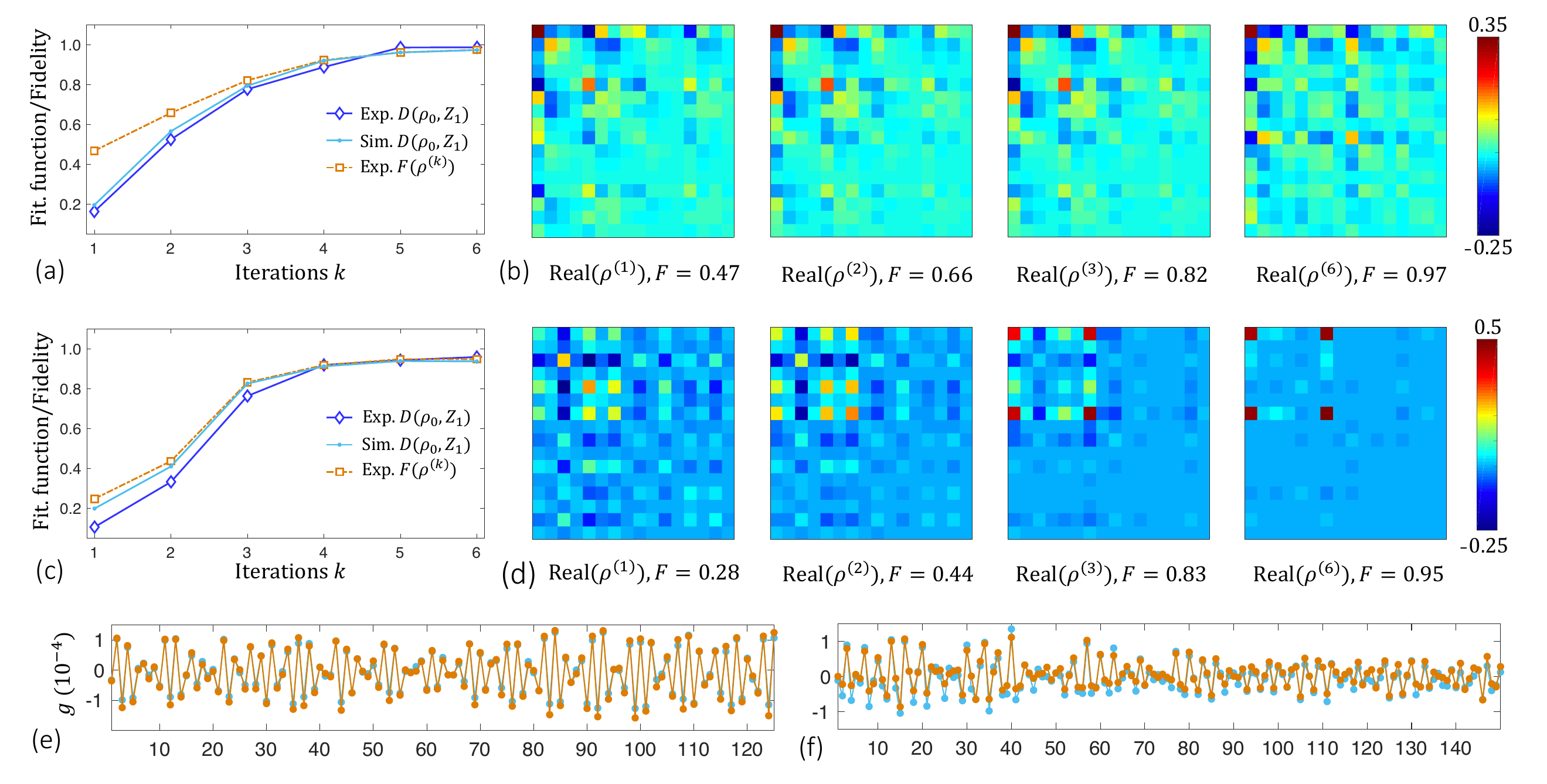}  
\caption{  {Experimental results in the reconstruction of $\rho$ by variational HQC-based QST. (a)-(b) The result for the dynamical state. (c)-(d) The result for the entangled state. (a) and (c) The fitness function and the fidelity as a function of iteration number $k$. The cyan and blue points represent   the simulated $D(\rho_0, Z_1)$ and the experimentally measured $D(\rho_0, Z_1)$    using the   pulse $b^{(k)}$ on the controlled system. $F(\rho^{(k)})$ is the fidelity of the reconstructed state $\rho^{(k)}$ with the   target state. (b) and (d) The real parts of the density matrix elements of the reconstructed $\rho^{(k)}$ for $k=1,2$, 3, and 6 for the dynamical state and the entangled state. They were estimated from $\mathcal{C}(b^{(k)})^{\dagger}\ket{0000}\bra{0000}\mathcal{C}(b^{(k)})$ by implementing the   pulse $b^{(k)}$ on the state $\ket{0000}$.  (e) and (f) The gradient vectors  $g^{(1)}_x$ for the case of the dynamical state and entangled state, respectively. The experimental and simulated data are labeled by the orange and the cyan, respectively.}} 
\label{results}
\end{figure*}

Last, we changed the control parameters $b^{(k)}_x[m]$ and $b^{(k)}_y[m]$ by a step size $\triangle=1$ kHz for the $m$-th slice to measure the gradient values $g^{(k)}_x[m]$ and $g^{(k)}_y[m]$. Here, (i) for the dynamical state, we set the number of slices as $M=125$ and the time length for each slice as $\tau=40$ $\mu$s for the initial pulse; (ii) for the entangled state, we set the number of slices as $M=150$ and the duration of each slice as $\tau=60$ $\mu$s. So there are in total $2\times 125=250$ control parameters to be optimized in each iteration for the dynamical state (300 control parameters for the case of entangled state).   After the gradient vector $g^{(k)}$   was determined in experiments, we updated the   pulse along the gradient direction for obtaining an increment of $D(\rho_0^{(k+1)}, Z_1)$.  Here, $2M$ experiments are necessary to determine the gradient values for one round of iteration.

After the optimization was finished, we also performed 4-qubit   QST on the final state to evaluate its fidelity with the ideal state  $\rho_{0}=\ket{0000}\bra{0000}$. We use  $\rho^{\text{e}}_{0}$ to denote the   experimentally reconstructed final state. We found that the fidelity between   $\rho^{\text{e}}_{0}$ and the ideal state $\rho_{0}$ is   0.986 for the dynamical state case and 0.943 for the entangled state case. This means that $\rho$ was almost driven into    $\ket{0000}\bra{0000}$ through the optimized pulse.

\subsection{Results} 
Now we present the results of the reconstruction via HQC-based QST for the dynamical state and the entangle state. 

  {\emph{The dynamical state}.-In this experiment, we in total performed 6 iterations and 250 experiments for each iteration such that the target state $\rho$ was reconstructed with   sufficient quality.  In Fig. \ref{results}(a), it is clear that the fitness function $D(\rho_0, Z_1)$ and the fidelity of the state $\rho^{(k)}$ are approaching the optimal value of 1 with the increasing number of iterations, and there is a good agreement between the experimentally measured and simulated $D(\rho_0, Z_1)$. Here, $\rho^{(k)}$ is the reconstructed state via our method at the $k$-th iteration, which is computed as $\mathcal{C}(b^{(k)})^{\dagger}\ket{0000}\bra{0000}\mathcal{C}(b^{(k)})$ via the   pulse $b^{(k)}$ on the NMR simulator. The real parts of the density matrices  $\rho^{(k)}$ for $k=1, 2, 3$, and 6 are presented in Fig. \ref{results}(b). Eventually, over 0.97 fidelity was achieved via our method in reconstructing the dynamical state. In each iteration, we measured the gradient vector $g^{(k)}$, whose precision determines the search direction of the optimization and hence the performance of the next iteration. Here, we place the comparison between the measured and simulated gradient vector $g^{(1)}_x$ in Fig. \ref{results}(e).}

\begin{figure}
\centering  
\includegraphics[width=0.85\linewidth]{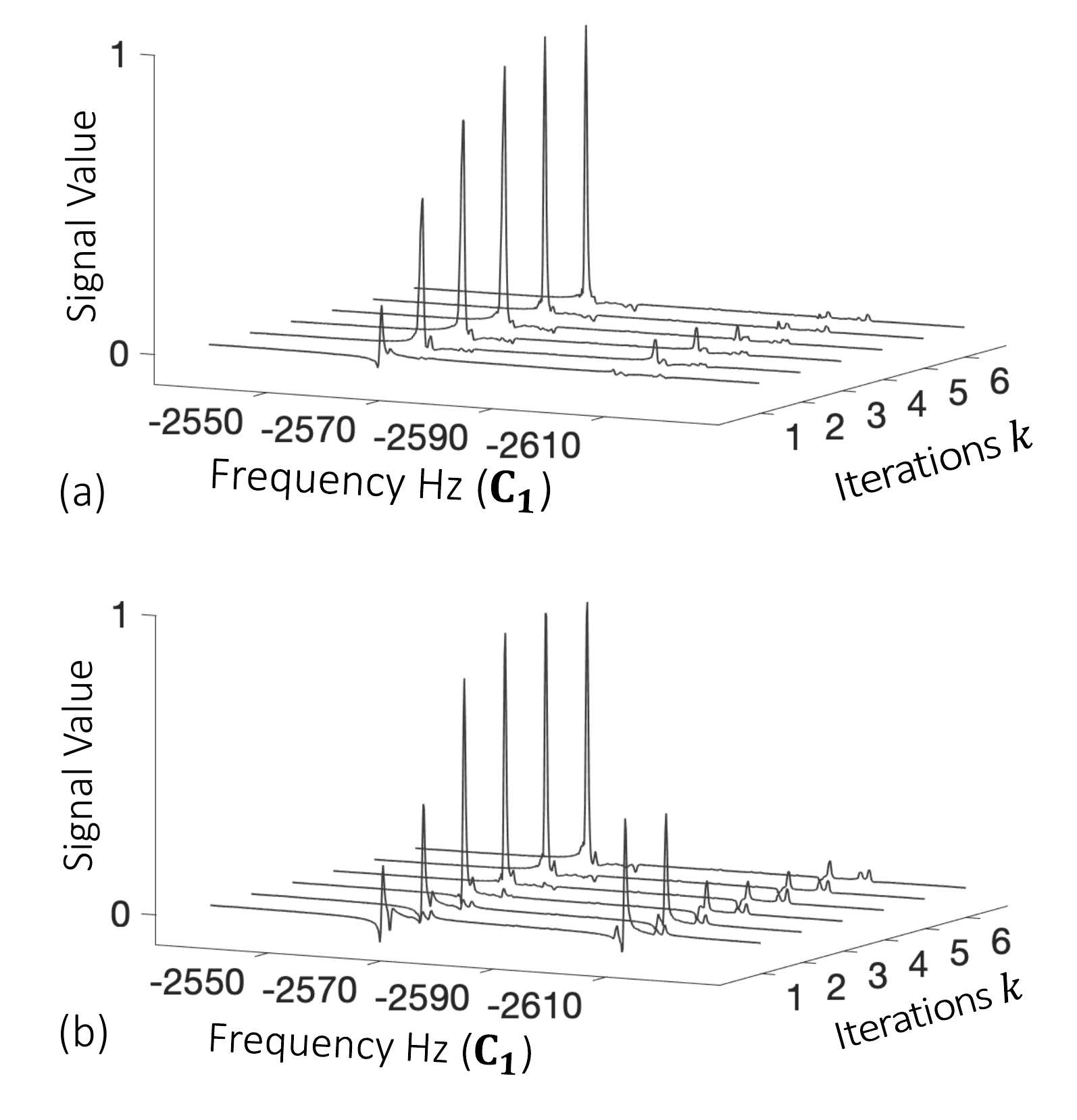}  
\caption{Experimental spectra of the spin C$_1$ for each iteration for the dynamical state (a) and the entangled state (b). The spectra were obtained by performing a $\pi/2$ readout pulse on the spin C$_1$ before data acquisition. Single-peak pattern appeared after $k=3$, and the intensity of this peak increased to about 0.9 after $k=4$. } 
\label{spectrum}
\end{figure}

\emph{The entangled state}.-The   results are shown in Fig. \ref{results}(c) and \ref{results}(d). Figure \ref{results}(c) presents the    simulated and the experimentally measured $D(\rho_0, Z_1)$, as well as the fidelity of $\rho^{(k)}$ as a function of the iteration number $k$. Here, the simulated $D(\rho_0, Z_1)$ was directly computed by applying the iterated pulse $b^{(k)}$ on the ideal state $\rho$ and numerically measuring the expectation value of $Z_1$ on the classical computer. $D(\rho_0, Z_1)$ and $D(\rho_0, \ket{0000}\bra{0000})$ can be both considered as the fitness function in this optimization. If $D(\rho_0, Z_1)$ converges to the optimal value, then so does the fitness function $D(\rho_0, \ket{0000}\bra{0000})$. As shown in Fig. \ref{results}(c), $D(\rho_0, Z_1)$ increased to 0.90 after 4 iterations, and then reached 0.95  when iteration number $k=6$. Similarly, we estimated the quality of the reconstructed state $\rho^{(k)}$ in each iteration, by applying the $k$th iterate pulse $b^{(k)}$ on the state $\ket{0000}\bra{0000}$. Figure \ref{results}(d) presents the real parts of the density matrices of the reconstructed states $\rho^{(k)}$ for iteration number $k=1, 2, 3$, and 6. It obviously shows that the state $\rho^{(k)}$ is approaching  the ideal state in experiments, and the density matrix form of Bell state between C$_2$ and C$_3$ shows up when $k=5$. Finally, we successfully realized the reconstruction of the entangled state via our method with around 0.95 fidelity. In each iteration, the gradient vector $g^{(k)}$ with 300 parameters was measured on the NMR simulator. Figure \ref{results}(f) presents the measured and simulated gradients for   the first iteration.

In the NMR spectrum, each qubit's signal  contains 8 peaks because of its coupling with the other three spins \cite{xin2017quantume}. 
%The real  and imaginary parts of the $i$-th peak represent the expectations of the operators $\sigma_x\otimes\ket{b(i-1)}\bra{b(i-1)}$ and $\sigma_y\otimes\ket{b(i-1)}\bra{b(i-1)}$, respectively. $b(i-1)$ is the binary number of $i-1$ in 3 bits, such as $\ket{b(2)}=\ket{011}$. 
If a $\pi/2$ readout pulse acting on the first spin is applied on the ideal state $\ket{0000}$, there should appear a single-peak signal labeled by the rest spins $\ket{000}$. 
%Hence, single-peak spectrum is usually the feature of the state $\ket{0000}$ in NMR (necessary but not sufficient condition). 
Figure \ref{spectrum} shows the spectra of C$_1$ as a function of iteration number $k$. Single-peak appearing after $k=4$ 
also implied that convergence was   almost achieved.

\subsection{Error Analysis and Convergence}
In experiments, there are certain error sources including the imperfections of  PPS initialization, the infidelities of the  GRAPE pulses, and decoherence effects. These error sources have consequences in the experimental results, which we describe as follows. (i) They cause a deviation between the experimental and the ideal results. Here, in order to estimate the influence of the potential errors from real experiments on the values $D(\rho_0, Z_1)$, we numerically simulated the quantum dynamics which starts from the prepared state in experiments and evolves under the GRAPE pulse with consideration of a decoherence model for each iteration. We further compared the simulated values $D_\text{sim}$ with the ideal ones $D_\text{th}$, and then computed the standard deviation of our simulated results as $\epsilon=\sqrt{\sum^K_i(D^i_\text{sim}-D^i_\text{th})^2/(K-1)}$, with $K$ the number of iterations. It is found that $\epsilon$ is 2.45\% for the dynamical state, and is 4.96\% for the entangled state. (ii) They lead to the inaccuracies in the measurement of the gradients. Imperfections in measuring gradients   deviate the search direction from the desired one and could cause the search be trapped by local extremum. (iii) %The experimental state $\rho^{\text{e}}$ is   only approximately pure  due to the presence of these errors. 
%It is impossible to definitely  optimize a unitary dynamics evolving this mixed state to a prefect state $\ket{0000}$. 
Minor experimental errors can lead to   inaccuracies of the gradients and the impurity of $\rho^e$. In that case, it is impossible to get a unitary sequence $\mathcal{C}(b)$ that can perfectly drive the mixed state $\rho^e$ to the pure state $\ket{\textbf{0}}\bra{\textbf{0}}$. This explains why our result in practice did not converge to the perfect fidelity of 1.

%\begin{figure}
%\centering  
%\includegraphics[width=0.85\linewidth]{hybrid_simm.pdf}  
%\caption{Iterative process when the target states are $\rho_{\text{tr}}$ (pure state) and $\rho^{\text{e}}_{\text{tr}}$ (mixed state) both from the initial guess pulse $b^{(1)}$.  The process starts from $\rho_{\text{tr}}$ and   eventually   converges to a value close to 1 after a number of iterations.  It actually only converges to about 0.92 due to the impurity of $\rho^{\text{e}}$. } 
%\label{simm}
%\end{figure}

\section{Conclusion}

We presented a  self-learning QST method via the variational HQC approach, in which the fitness function and gradients are measured online with the  system itself. Hence, such a method can reconstruct the real states in experiments with the advantages of  reliability and efficiency. While the conventional brute-force full state tomography approach needs to perform exponential number of measurements, our method is promising for provide a better scaling for a broad class of practical quantum states, e.g., dynamical states of a quantum many-body system. Another advantage is that,  unlike   conventional QST methods that are based on   raw experimental measurement data, which usually output   unphysical density matrices (e.g., not positive semi-definite) and thus require some subsequent correction techniques like maximum likelihood estimation \cite{singh2016constructing},  the HQC method directly outputs the valid density matrix once the desired control sequence is found. Therefore, our work offers a new way to enhance the efficiency of   QST  in practice.

  {The variational HQC-based QST  method   can be applied to but not limited to  reconstruction of the dynamical states of quantum many-body systems, the ground states of $k$-local Hamiltonians in quantum statistical mechanics, and the entangled quantum states commonly used in quantum computing and quantum communication. Undoubtedly, these states are frequently encountered  in quantum experiments and are  of importance for emerging quantum  technologies. Many recent known research work have been devoted to the tomography  problem for these quantum states \cite{bairey2019learning,xin2018local,torlai2019integrating,Lanyon17,Torlai18,Ohliger_2013}. Our variational HQC method offers an alternative choice.  Using techniques of NMR, we have experimentally demonstrated  the feasibility of performing QST via our method by successfully reconstructing a 4-qubit dynamical state of quantum many-body system and an entangled state.   For future studies, it would be interesting to extend the  current method   to other experimental platforms. We expect the methodology developed here can become a useful tool for  practical quantum tomography on   intermediate-scale quantum   devices that are about to appear in the near term.}

\begin{acknowledgments}
T. X., J. L. and D. L. are supported by the National Key Research and Development Program of China (Grants No. 2019YFA0308100), National Natural Science Foundation of China (Grants No. 11605005, No. 11875159, No. U1801661, No. 11905099, and No. 11975117), Science, Technology and Innovation Commission of Shenzhen Municipality (Grants No. ZDSYS20170303165926217 and No. JCYJ20170412152620376), Guangdong Innovative and Entrepreneurial Research Team Program (Grant No. 2016ZT06D348), Guangdong Basic and Applied Basic Research Foundation (Grant No. 2019A1515011383).

\end{acknowledgments}

%\bibliography{tomo.bib}


\end{document}